\documentclass[superscriptaddress,twocolumn,showpacs,preprintnumbers,amsmath,amssymb, prb, reprint, longbibliography]{revtex4-1}
\usepackage{amsmath}
\usepackage{amssymb}
\usepackage{graphicx}
\usepackage{bm}
\usepackage{epsfig}
\usepackage{color}

\usepackage[colorlinks=true, linkcolor=blue, citecolor=blue, urlcolor=blue]{hyperref}
\begin{document}

\title{Distinguishing between $s+id$ and $s+is$ pairing symmetries in multiband superconductors through spontaneous magnetization pattern induced by a defect}
\author{Shi-Zeng Lin}
\affiliation{Theoretical Division, Los Alamos National Laboratory, Los Alamos, New Mexico 87545, USA}
\author{Saurabh Maiti}
\affiliation{Department of Physics, University of Florida, Gainesville, FL-32611, USA}
\author{Andrey Chubukov}
\affiliation{Department of Physics, University of Minnesota, Minneapolis, Minnesota 55455, USA}

\begin{abstract}
The symmetry of the pairing state in iron pnictide superconductor $\mathrm{Ba_{1-x}K_xFe_2As_2}$ is still controversial. At optimal doping ($x \approx 0.4$), it is very likely $s$-wave, but for $x=1$ there are experimental and theoretical arguments for both $s$-wave and $d$-wave. Depending on the choice for $x=1$, intermediate $s+is$ and $s+id$ states have been proposed for intermediate doping $ 0.4 < x < 1$. In both states, the time reversal symmetry is broken and a spontaneous magnetization is allowed. In this work we study a spontaneous magnetization induced by a nonmagnetic defect in the $s+is$ and $s+id$ states by using a perturbation theory and  numerical calculations for the Ginzburg-Landau free energy functional. We show that  the angular dependence of the magnetization is distinct in these two states due to the difference in symmetry properties of the order parameters. Our results indicate a possible way to distinguish between the $s+is$ and $s+id$ pairing symmetries in multi-band superconductors.

\end{abstract}
\date{\today}
\maketitle

\section{Introduction}
  The knowledge of the  pairing symmetry of an unconventional superconductor is the first step to
   elucidate the pairing mechanism.
  In iron pnictide and iron-chalcogenide superconductors (FeSCs) the identification of the pairing symmetry is complicated
   by the following  reasons. First, the family of iron-based superconductors is large, and it is not certain that the pairing symmetry is the same in all FeSCs.
   The two major candidates are $s_{+-}$ \cite{Mazin08,Kueoki08,PhysRevB.78.134524,PhysRevB.78.134512,PhysRevLett.101.206404} and
    $d_{x^2-y^2}$ \cite{Kueoki08,PhysRevLett.101.206404,PhysRevLett.102.047006,Thomale} states.  At optimal doping $s_{+-}$ state is favorable, but at larger hole doping
     several theoretical work suggested that $s_{+-}$ and $d_{x^2-y^2}$  are
     almost degenerate  \cite{1367-2630-11-2-025016,Maiti_10,Thomale}. Second, FeSCs are multi-band superconductors with several hole and electron
      Fermi pockets.  The phase of $s_{+-}$ gap changes by $\pi$ between hole and electron pockets and the phase of $d_{x^2-y^2}$ gap changes by $\pi$ between electron pockets.  In general, however, multi-band superconductors allow more complex superconducting states with the phase differences of the gaps on different Fermi pockets as fractions of $\pi$.

These more complex superconducting states, even with pure $s-$wave symmetry, can be understood by noticing that
the $s_{+-}$ pairing symmetry originates from an
interband repulsive interaction between superconducting condensates on hole and electron pockets. The intra-pocket repulsion
favors a $\pi$ phase shift in the superconducting order parameter.
When three
 or more
 bands
  are present, the inter-band repulsion leads to frustration, which is resolved by  choosing the phases of the gaps to maximize superconducting condensation energy.
   For example, for three identical pockets with equal repulsive inter-band interaction, the best outcome is the $\pm 2\pi/3$ difference between the phases of the three gaps, much like
    $120^{\circ}$ spin configuration resolves frustration in an XY antiferromagnet on a triangular lattice. And just like there, the phase change can go in $2\pi/3$ ingredients clockwise or anticlockwise. The choice
 breaks the time reversal symmetry in addition to the overall phase $U(1)$ symmetry. \cite{Agterberg99,Stanev10,Hu11,PhysRevB.88.220511}
  In a generic state of this kind, the order parameter $\hat{\Psi}\equiv (\Psi_1,\ \Psi_2,\ \Psi_3, ...)$ made out of
   $\Psi_i\equiv \Delta_i\exp(i\varphi_i)$ at a given Fermi pocket, and its complex conjugate $\hat{\Psi}^*\equiv (\Psi^*_1,\ \Psi^*_2,\ \Psi^*_3, ...)$
    are not related by a global phase rotation, i.e.,  $\hat{\Psi} \neq \exp(i\theta_0)\hat{\Psi}^*$.  Below we label such $s$-wave state as $s+is$.

    The transition from a time-symmetry preserving  $s$-wave state (e.g., $s_{+-}$) to an $s+is$ state
  is a continuous phase transition. Near the transition, a collective out-of-phase oscillation of $\varphi_i$ (a Leggett mode) becomes soft. \cite{Lin2012PRL,Stanev11,LinReviewMultibandSC} Inside an $s+is$ state, the breaking of a discrete time reversal symmetry  gives rise to  a new kind of a phase soliton between the superconducting domains $\hat{\Psi}$ and $\hat{\Psi}^*$.~\cite{PhysRevLett.107.197001,LinNJP2012,PhysRevLett.112.017003} It has been proposed that one may detect the $s+is$ pairing symmetry based  these properties. \cite{Lin2012PRL,LinNJP2012,PhysRevLett.112.017003,Marciani2013,Burnell10,Maiti2013}

FeSCs are promising candidates for an $s+is$ state. The material $\mathrm{Ba_{1-x}K_xFe_2As_2}$ is particularly interesting in this regard. It is magnetic at small $x$ and superconducting at larger $x$.
  Its electronic structure for $x$ not close to $x=1$ consists of three hole pockets and two electron pockets. Near optimal doping, it is very likely that superconducting interaction is mediated by spin fluctuations. The
   magnetic order has the momentum equal to the distance between the centra of
   hole and of electron pockets, and fluctuations of this order clearly favor $s_{+-}$ state without breaking the time-reversal symmetry and  $\pi$ phase shift between the gaps on  hole and electron pockets.  This is consistent with angle-resolved photoemission spectroscopy (ARPES) experiments, which near $x=0.4$ found a fully gapped superconducting state with little variation of the gaps along the pockets.
  \cite{Ding08,PhysRevB.83.020501,PhysRevLett.102.187005,PhysRevB.80.140503,christianson_unconventional_2008}.  At $x=1$, however, the  situation is different: electron pockets disappear and only the hole pockets remain, according to the ARPES measurements. \cite{PhysRevLett.103.047002,okazaki_octet-line_2012} If the pairing symmetry remains $s$-wave all the way to $x=1$, as  some ARPES experiments suggested based on the measurements on the gaps on hole pockets~\cite{okazaki_octet-line_2012},  then the gap must change sign between inner and middle hole pockets, where the gaps are the largest
    \cite{okazaki_octet-line_2012}. The sign-changing $s$-wave gap structure was found in random phase approximation based theoretical studies for $x=1$~\cite{Maiti_10}.   If one analyses how to connect $s_{+-}$ state near optimal doping, with equal sign of the gaps on the hole pockets, and $s$-wave state at $x=1$ with opposite signs of the gaps on the two smallest hole pockets, one
     finds~\cite{Maiti2013,Marciani2013} that the evolution is continuous near the $T_c$ line, with the gap on one of hole pockets vanishing and re-appearing with a different sign as $x$ approaches 1,  but necessary  involves an intermediate state at $T=0$, when it is easier to change the phase of the order parameter rather than its amplitude. In this intermediate state,  the phases of the gaps on the two smallest hole pockets differ by a fraction of $\pi$, i.e.,  the intermediate state is a realization of $s+is$ superconductivity.

    Another suggestion, based on measurements of thermal conductivity and resistivity~ \cite{reid_d-wave_2012,PhysRevLett.109.087001,tafti_sudden_2013,PhysRevB.89.134502} and theoretical calculations using functional renormalization group \cite{Thomale}, is that
     the pairing symmetry at $x=1$ is a $d$-wave ($d_{x^2-y^2}$).  This is also generally consistent with random phase approximation based calculations~\cite{Maiti_10}, which found that $s$-wave and $d_{x^2-y^2}$ couplings are almost degenerate.  If so, the system must evolve from an $s_{+-}$ superconductor at $x \sim 0.4$ to a $d-$wave superconductor at $x=1$.
      Calculations show \cite{PhysRevLett.106.187003,1367-2630-11-5-055058} that this evolution goes via an intermediate phase in which
      both  $s$-wave and $d$-wave components are present, and the phase shift between the two is  $ \pm \pi/2$, i.e, the state is $s + id$.
        This is another state which breaks time-reversal pairing symmetry.

 In this work we discuss whether it is possible to distinguish between  $s+is$  and $s+id$  states in experiments.
  Both states break  time reversal symmetry and allow a spontaneous magnetization to develop. In a homogenous system,  magnetization does not develop
  because  a spontaneous current circulates in the
   band space and is not coupled to a gauge field.
     However, in the presence of nonmagnetic defects,
      a spontaneous current can be induced around the defect in superconductors with either $s+is$ or $s+id$ pairing symmetries \cite{PhysRevLett.102.217002,LinNJP2012,PhysRevB.91.161102,PhysRevLett.112.017003}. Because the $s+id$ state breaks the lattice $C_4$ rotation symmetry, while the $s+is$ preserves it, the profile of the induced spontaneous magnetization are different for  $s+is$ and $s+id$ states. This suggests a possible way to experimentally distinguish between the two pairing symmetries.

  Below we report the results of our study on the spontaneous magnetization in $s+is$ and $s+id$ superconductors induced by  nonmagnetic defects with different shapes.
   We extend the previous work \cite{PhysRevLett.102.217002,PhysRevB.91.161102} by developing a self consistent  treatment for the magnetization based on a phenomenological Ginzburg-Landau free energy functional. Our results show that one can differentiate between the $s+is$ and $s+id$ states by measuring the magnetization pattern induced by nonmagnetic defects. We  first present the symmetry argument to obtain the magnetization profile. We  then present perturbative calculations of  induced magnetization for a weak defect potential.  We next compare the perturbative calculations  with the numerical results obtained by minimizing the Ginzburg-Landau free energy functional. Finally, we  compare our approach and the results to those in  previous work.

\section{Model and perturbative calculations}

\subsection{Symmetry analysis}

\begin{figure}[t]
\psfig{figure=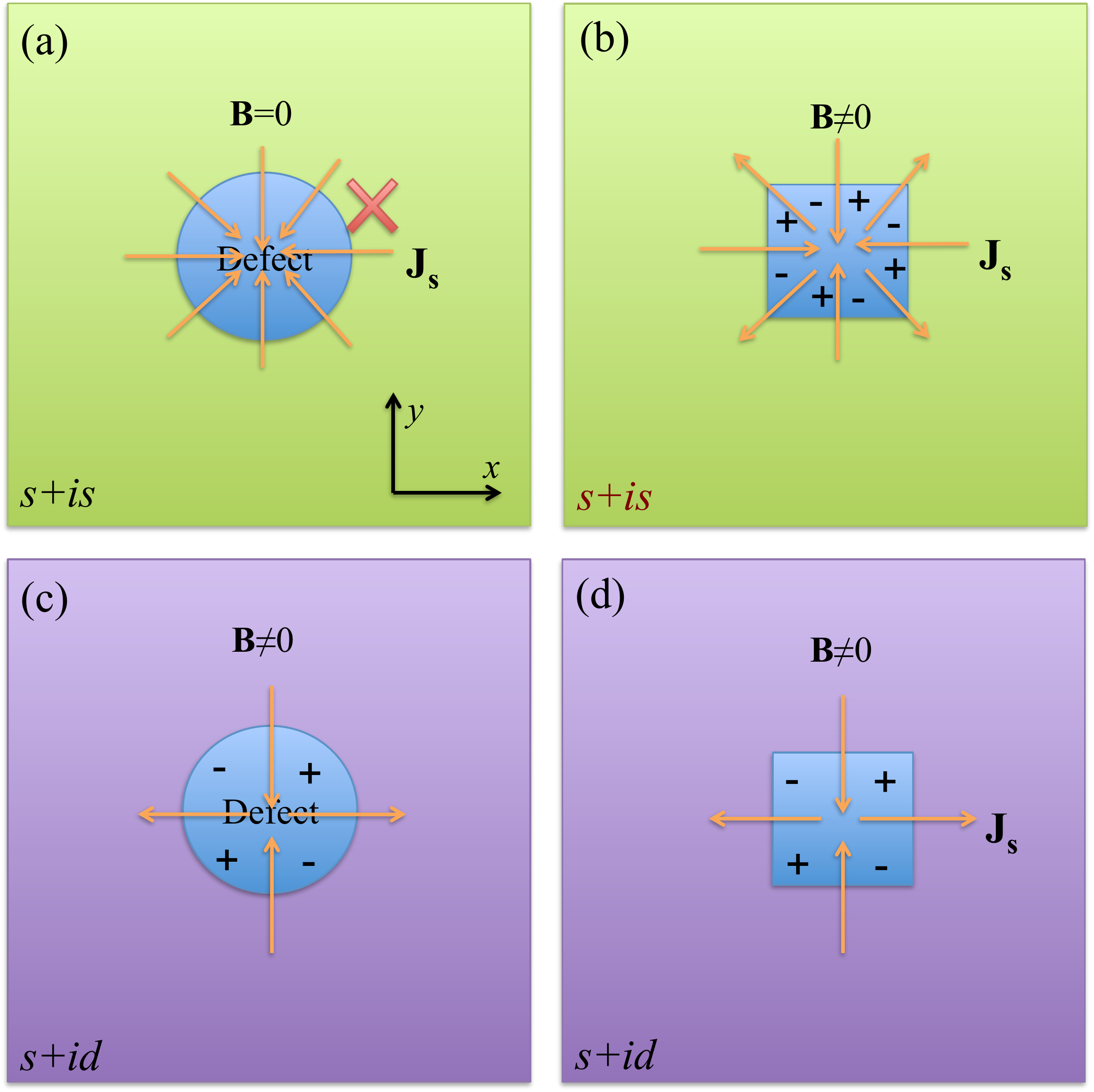,width=\columnwidth}
\caption{(color online) Schematic view of the spontaneous supercurrent and magnetization induced by a circular and square defect in the $s+is$ and $s+id$ states based on the symmetry analysis. The arrows are the direction of supercurrent and $+(-)$ denotes the direction of the magnetization field perpendicular to the superconducting plane.
} \label{f1}
\end{figure}

Before going into  detailed calculations, let us first perform a symmetry analysis. The $s+is$ state has a full rotational symmetry.
For a circular defect, if a supercurrent was
 induced,
  it could only
  flow inward or outward, as sketched in Fig. \ref{f1} (a). This would violate the current conservation $\nabla\cdot\mathbf{J}_s=0$, hence
   no supercurrent (and no spontaneous magnetization) is allowed in this case.
    Consider next
     a square defect.
        Because $x$ and $y$ directions are equivalent, to conserve the supercurrent, the direction of the supercurrent
         along zone diagonals
         should be opposite to that in the $x$ and the $y$ directions. That is if the current flows inward in the $x$ and the $y$ directions, the current must flow outward in the diagonal direction, and vice versa.
         As a result,
          the induced magnetization, measured as a function of the angle with respect to, say, $x$ direction,
            must display
             a four-fold oscillation
              [Fig. \ref{f1} (b)].

              The
              $s+id$ state breaks the $C_4$ rotation symmetry and
              is invariant under the combination of the $C_4$ rotation and time reversal operation.
              In this situation, a supercurrent and a spontaneous magnetization do emerge, even if a defect is circular.
               For a circular or square defect, if the current in the $x$ direction flows inward, then the current in the $y$ direction must flow outward, and vice versa. The current conservation at the center of a  defect
                is satisfied automatically. The
                magnetization pattern, generated by a supercurrent,
                 displays a two-fold oscillation as a function of an angle [Fig. \ref{f1} (c) and (d)].
                 This simple analysis shows that
                  one can indeed differentiate between an $s+is$ pairing state and an
                   $s+id$ pairing
                     state by analyzing the pattern of a
                     spontaneous magnetization induced around a defect.

\subsection{Ginzburg-Landau free energy}

We next calculate a spontaneous magnetization induced by a defect using Ginzburg-Landau theory in two dimensions.  We consider $s+is$ and $s+id$ states separately.

\subsubsection{$s+is$ state}

Like we said,
  an $s+is$ state emerges when there are three (or more) Fermi pockets, due to
  frustration when inter-pocket interactions  are repulsive.
    \cite{Stanev10,Marciani2013,Maiti2013}  However, recent work  \cite{Garaud2016} has demonstrated that one
     can simplify the analysis of $s+is$ state by
       reducing the three-pocket model to an effective two-pocket model in which
           time reversal symmetry breaking is explicitly imposed.
            In this approach, which we follow,
            the
            Ginzburg-Landau free energy functional for an $s+is$ state is, up to terms of  quartic order in $\Psi_i$:
\begin{equation}\label{eq1}
\mathcal{F}\left(\Psi _i, \Psi _i^*\right)=\mathcal{F}_1+\mathcal{F}_2+\mathcal{F}_c+\frac{1}{8 \pi }(\nabla \times \mathbf{A})^2,
\end{equation}
where the free energy density for each component $\Psi_i$ is ($i =1,2$)
\begin{equation}\label{eq2}
\mathcal{F}_i= {{\alpha _i}|{\Psi _i}{|^2} + \frac{{{\beta _i}}}{2}|{\Psi _i}{|^4} + \frac{\hbar^2}{{2{m_i}}}{{\left| {\hat{\mathbf{D}}{\Psi _i}} \right|}^2}},
\end{equation}
and the  coupling between $\Psi_1$ and $\Psi_2$ is described by
\begin{align}\label{eq3}
{{\cal F}_{c}} = \frac{{{\gamma _3}}}{2}|{\Psi _1}{|^2}|{\Psi _2}{|^2} + \frac{{{\gamma _2}}}{2}\left[ {{{\left( {\Psi _1^*{\Psi _2}} \right)}^2} + c.c.} \right]  \nonumber\\
+ {\gamma _1}\left( {{\Psi _1}\Psi _2^* + c.c.} \right)+ \frac{{{\hbar ^2}}}{{4{m_c}}}\left[ {{{\left( {\hat{\mathbf{D}}{\Psi _1}} \right)}^*}\cdot\hat{\mathbf{D}}{\Psi _2}  + c.c.} \right],
\end{align}
where $\hat{\mathbf{D}}=-i\nabla-2\pi \mathbf{A}/\Phi_0$ and $\Phi_0=hc/2e$ is the flux quantum. Because of competition between bilinear and bi-quadratic terms (the $\gamma_1$ and the $\gamma_2$ terms), the phase difference between $\Psi_1$ and $\Psi_2$ can be any value and the resulting state generally can be termed as $s+\exp(i\varphi) s$ state [for analogous consideration in coexistence state of superconductivity and magnetism, see Ref. \onlinecite{alb}].  To stabilize the $s+is$ state, we set $\gamma_2>0$ and choose the phase shift between $\Psi_1$ and $\Psi_2$ to be $\varphi=+\pi/2$. The coupling between the bands at the bilinear level is via the $\gamma_1$ and $1/m_c$ terms. The $\gamma_1$ term may be safely set to zero, as in any case it can
  be eliminated by an appropriate rotation in $(\Psi_{1}, \Psi_2)$ space.  This procedure changes the  values of $\gamma_{2,3},~1/m_i$ and $1/m_c$, but does not introduce new terms.
   For $s+id$ case the bilinear terms are not allowed by symmetry.

A nonmagnetic defect is modeled by changing $\alpha_i\rightarrow \alpha_i+\tilde{\alpha}_i (r)$. A defect is considered as weak if  the defect potential
$\tilde{\alpha}_i\ll \alpha_i$. The variation of the order parameter $\Psi_i (r)=\Psi_{i0}[1+\delta_i (r)+i \phi_i(r)]$, where  $\Psi_{i0}$ is the order parameters in the absence of defects, can be found from the
minimization of the Ginzburg-Landau functional to linear order in $\tilde{\alpha}_i (r)$. The minimization obviously yields  $\delta_i (r)$, $\phi_i (r) \propto \tilde{\alpha}_i (r)$. In a state with broken time-reversal symmetry an amplitude fluctuation $\delta_i (r)$ and a phase fluctuation $\phi_i (r)$ are coupled. As the consequence, a vector potential $\mathbf{A} (r)$ also fluctuates. Fluctuations of $\mathbf{A} (r)$ are gapped by Anderson-Higgs mechanism with the gap larger than that of collective excitations of $\delta_i (r)$ and $\phi_i(r)$, at least near the onset of a state which breaks time reversal symmetry.~\cite{Lin2012PRL,Stanev11} To calculate $\delta_i (r)$ and $\phi_i (r)$ to the linear order in $\tilde{\alpha}_i (r)$, we fix the gauge by choosing the global phases such that $\bf{A}=0$ when $\tilde{\alpha}(\bf{r})=0$. The resulting equations for $\delta_i$ and $\phi_i$ are presented in the Appendix. We assume that the  defect potential $\tilde{\alpha}_i(\mathbf{r})$ has the angular dependence in the form $\tilde{\alpha}_i(\mathbf{r})=\tilde{\alpha}_i(r)\cos(n\theta)$ and consider different integer $n$. The angular dependence  of $\delta_i (r)$ and $\phi_i (r)$  follows that of $\tilde{\alpha}_i(\mathbf{r})$.

Minimizing next $\mathcal{F}$ with respect to $\mathbf{A}$, we obtain the Ampere's law
\begin{equation}\label{eq4}
\nabla\times\nabla\times\mathbf{A}=\frac{4\pi}{c}\mathbf{J}_s,
\end{equation}
with the supercurrent
\begin{align}\label{eq5}
&\mathbf{J}_s=\sum_i \frac{{2e\hbar }}{{{m_i}}}\left| {\Psi _{i0}^2} \right|\left( {\nabla {\phi _i} + 2{\delta _i}\nabla {\phi _i}} \right) - \frac{{4{e^2}}}{c}\sum_i \frac{{\left| {\Psi _{i0}^2} \right|}}{{{m_i}}}{\bf{A}} + \nonumber \\
& {\frac{{\hbar e}}{{{m_c}}}{\left|\Psi_{10}\Psi_{20}\right|}\left( {\nabla {\delta _2} + {\delta _1}\nabla {\delta _2} + {\phi _1}\nabla {\phi _2} - \nabla {\delta _1} - {\delta _2}\nabla {\delta _1} - {\phi _2}\nabla {\phi _1}} \right)} .
\end{align}
The condition for current conservation, $\nabla\cdot \mathbf{J}_s=0$, is satisfied automatically by Eq. \eqref{eq4}.
Taking $\nabla\times$ of both sides of Eq. \eqref{eq4} and using $\nabla\times\nabla\phi_i=0$, which is  valid
as long as no vortices are present, we obtain
\begin{align}\label{eq6}
- {\nabla ^2}{\bf{B}} + {\lambda ^{ - 2}}{\bf{B}} =  \frac{{16\pi \hbar e}}{c}\sum_i \frac{\nabla {\delta _i} \times \nabla {\phi _i} }{{{m_i}}}\left| {\Psi _{i0}^2} \right| \nonumber \\
+ \frac{{8\pi \hbar e \left|\Psi_{10}\Psi_{20}\right|}}{c m_c}({ {\nabla {\delta _1} \times \nabla {\delta _2} + \nabla {\phi _1} \times \nabla {\phi _2}}}) ,
\end{align}
where $\lambda^{-2}=16\pi e^2\sum_i \left| {\Psi _{i0}^2} \right|/(m_ic^2)$ is the London penetration depth (up to
corrections of order $\delta_i$). The induced magnetization is of second order in the defect potential $\tilde{\alpha} (r)$ and it is directed
 perpendicular the 2D superconductor plane.

For a model of a superconductor with identical  $ \Psi_{10} = \Psi_{20}$ and $m_1=m_2$, $\delta_1 (r)=\delta_2 (r)$, and $\phi_1 (r)=-\phi_2 (r)$, the magnetization is absent because the right-hand side of Eq. (\ref{eq6}) vanishes.
   In a more generic case,  using $\delta_i(\mathbf{r})=\delta_i(r)\cos(n\theta)$ and $\phi_i(\mathbf{r})=\phi_i(r)\cos(n\theta)$, we obtain the magnetization field
\begin{align}\label{eq7}
\mathbf{B}(r, \theta)=\hat{z}\sin(2n\theta)\int_0^{\infty} B(k)J_{2n}(k r) k dk,
\end{align}
 where
\begin{align}\label{eq8}
B(k)=\frac{8\pi\hbar e n}{c r (k^2+\lambda^{-2})}S(k),
\end{align}
and
\begin{align}\label{eq9}
S(k)=\int_0^{\infty}  r drJ_{2n}(kr)\left[\sum_i \frac{{2\left| {\Psi _{i0}^2} \right|\left( {{\delta _i}{\partial _r}{\phi _i} - {\phi _i}{\partial _r}{\delta _i}} \right)}}{{{m_i}}} \right. \nonumber\\
\left. +\frac{{{\delta _1}{\partial _r}{\delta _2} - {\delta _2}{\partial _r}{\delta _1} + {\phi _1}{\partial _r}{\phi _2} - {\phi _2}{\partial _r}{\phi _1}}}{{{m_c}}}\left|\Psi_{10}\Psi_{20}\right|\right].
\end{align}
The unit vector $\hat{z}$ is along the $z$ direction, and $J_{2n}$ is the Bessel function of the first kind.
  We see that a
   nonmagnetic defect with an $n$-fold angular
     variation $\tilde{\alpha}_i (r) =\tilde{\alpha}\cos(n\theta)$
       induces a magnetization which displays a $2n$-fold oscillation:  $B_z\propto \tilde{\alpha}^2\sin(2n\theta)$. For a centrosymmetric defect potential $n=0$, the induced magnetic field vanishes.

\subsubsection{$s+id$ state}

We proceed to study the magnetization pattern from a defect
in an $s+id$ state.
  Let's denote an $s$-wave component  by $\Psi_1$ and a $d$-wave component by $\Psi_2$. The Ginzburg-Landau free energy functional is still given by Eqs. \eqref{eq1} and \eqref{eq2},
   but
    the coupling term is different and is given by
    \cite{PhysRevLett.102.217002}
\begin{align}\label{eq10}
\mathcal{F}_c=\frac{{{\gamma _3}}}{2}|{\Psi _1}{|^2}|{\Psi _2}{|^2} + \frac{\gamma_2 }{2}\left[ {{{\left( {\Psi _1^*{\Psi _2}} \right)}^2} + c.c.} \right]\nonumber\\
+\frac{{{\hbar ^2}}}{{4{m_c}}}\left[ {{{\left( {{\hat{D}_x}{\Psi _1}} \right)}^*}{\hat{D}_x}{\Psi _2} - {{\left( {{\hat{D}_y}{\Psi _1}} \right)}^*}{\hat{D}_y}{\Psi _2} + c.c.} \right].
\end{align}
For the $s+id$ state, $\mathcal{F}$ is invariant under a rotation of a reference frame by $\pi/2$ and subsequent the time reversal operation, as both operations change the sign of a $d$-wave component $\Psi_2$.
 The $\gamma_1$ term in Eq. \eqref{eq3} is not invariant under the change of the sign of $\Psi_2$ and is not allowed.
  Note also that the mixed gradient terms in the $x$ and the $y$ directions must have different signs.
 We show below that, because of the mixed gradient term in Eq. \eqref{eq10}, the magnetization is nonzero already to linear order in a defect potential.

\begin{figure}[t]
\psfig{figure=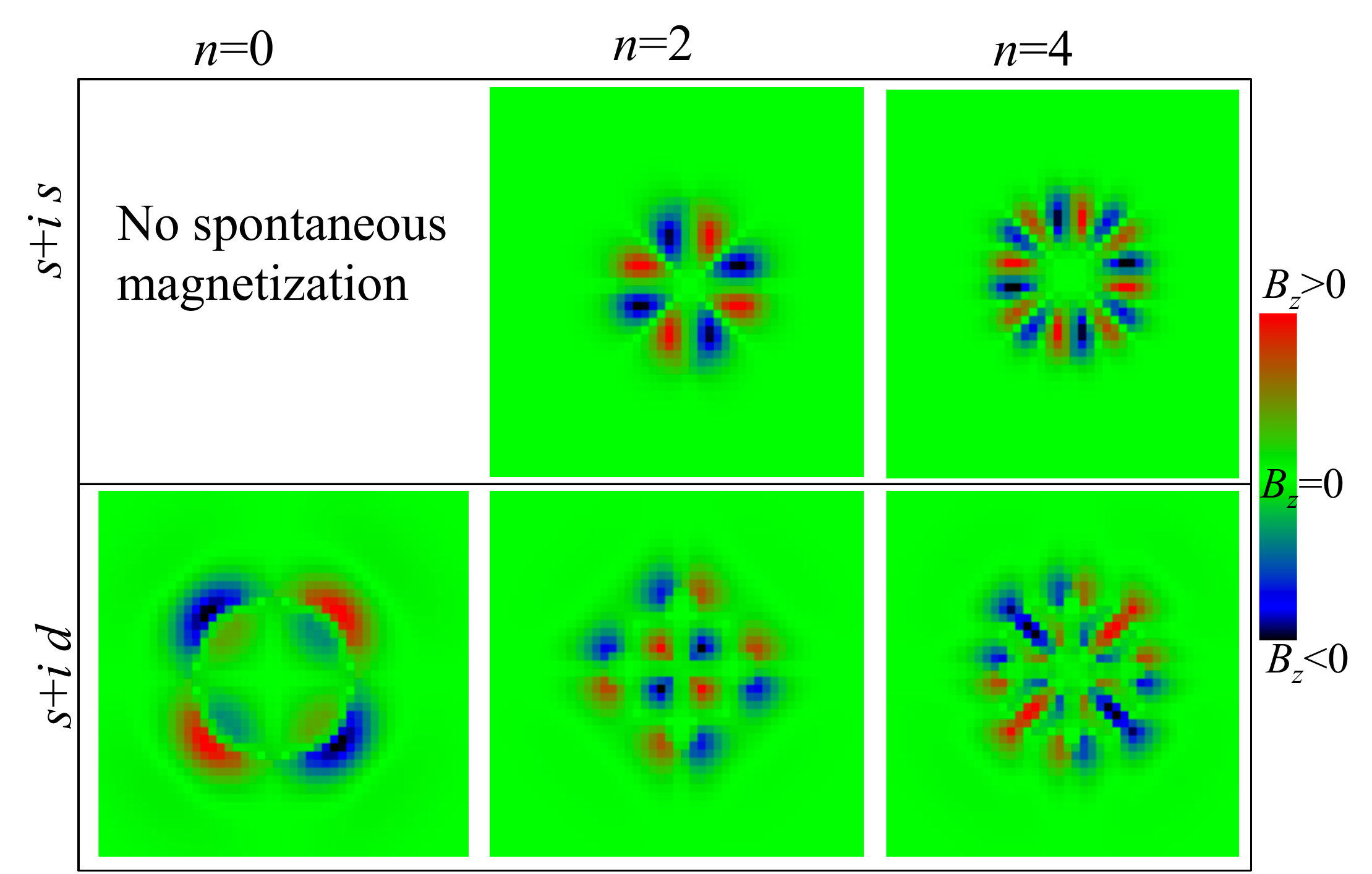,width=\columnwidth}
\caption{(color online) Spontaneous magnetization induced by a weak defect with $n$-fold angular dependence in the $s+is$ and $s+id$ states, obtained by numerical minimization of the Ginzburg-Landau free energy functional. The magnetization only has the component perpendicular to the superconducting plane. Here $\alpha_0=-1$, $\beta_i=1$, $\gamma_2=0.5$, $m_1=m_c=1$ and $m_2=2$. We consider a weak defect with $\tilde{\alpha}_i=0.1$ and $r_0=1.5$.} \label{f2}
\end{figure}

\begin{figure}[b]
\psfig{figure=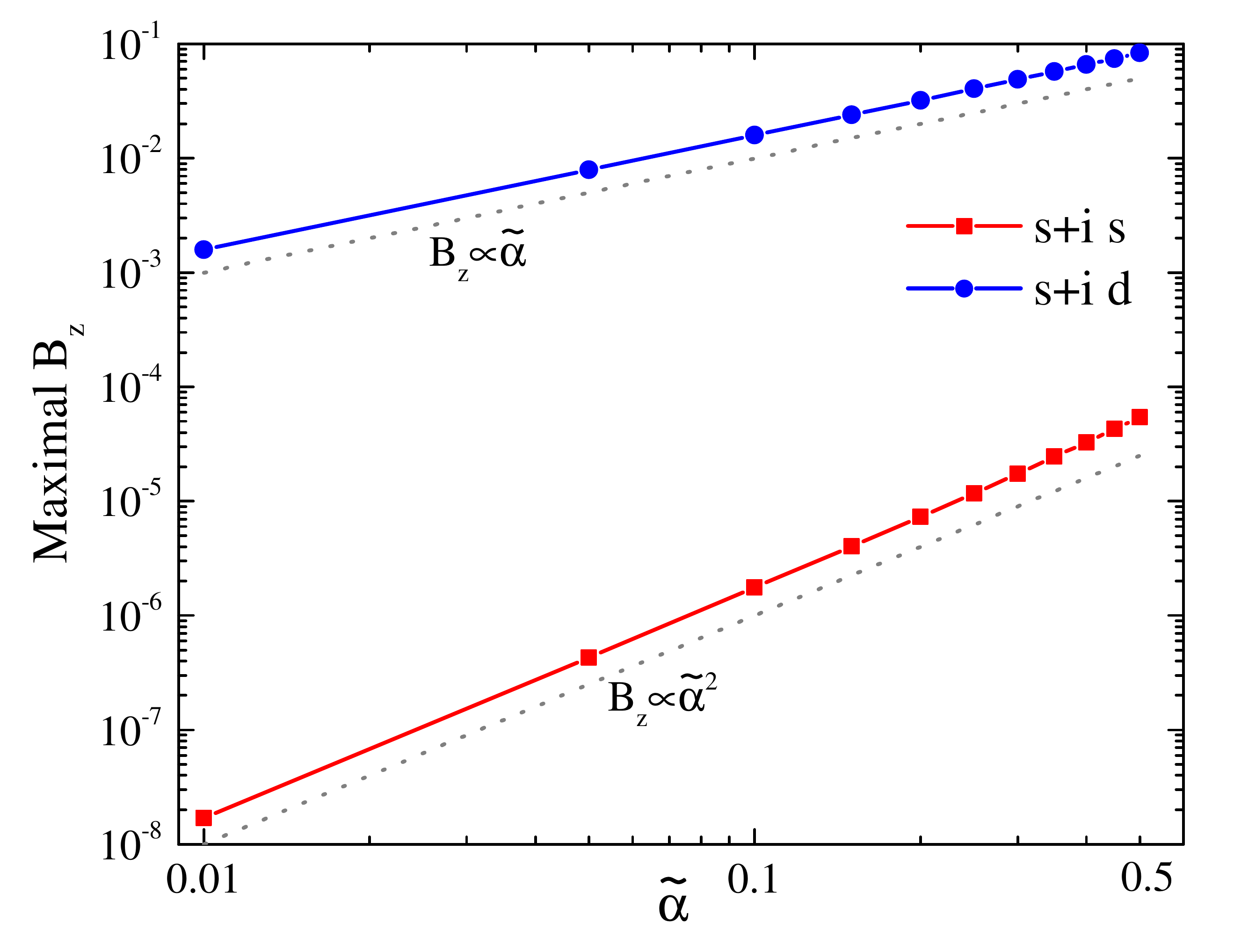,width=\columnwidth}
\caption{(color online) Maximal magnetization field induced by a square defect in the $s+is$ and $s+id$ states. Symbols are numerical results and lines are guide to eyes. The parameters are the same as those in Fig. \ref{f2} but with a different strength of the defect potential.
} \label{f3}
\end{figure}

\begin{figure}[t]
\psfig{figure=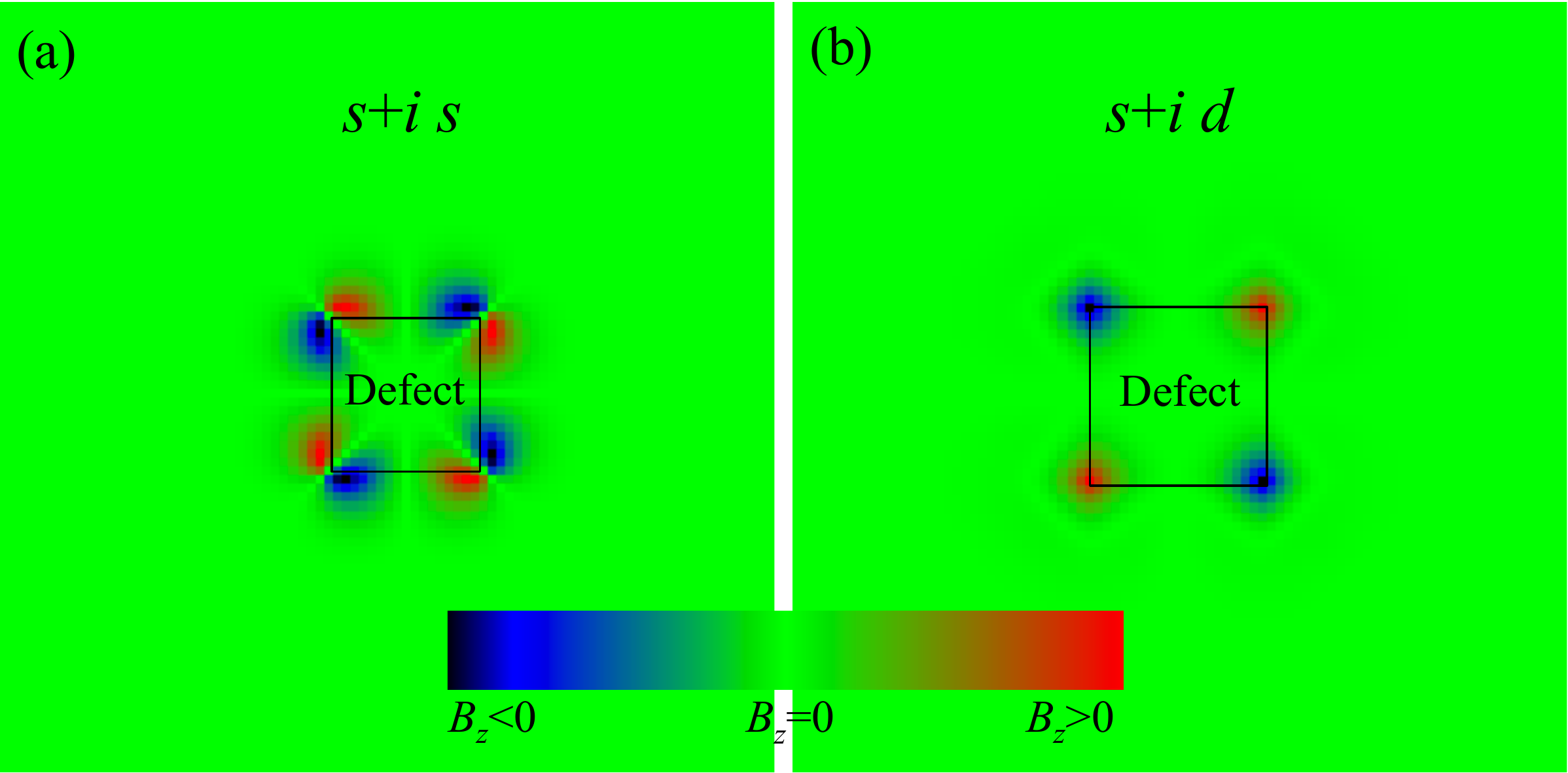,width=\columnwidth}
\caption{(color online) Spontaneous magnetization induced by a square defect in the $s+is$ and $s+id$ states. The magnetization only has the component perpendicular to the superconducting plane. Here $\alpha_0=-1$, $\beta_i=1$, $\gamma_2=0.5$, $m_1=m_c=1$ and $m_2=2$. The strength of the defect potential is $\tilde{\alpha}_i=0.5$ and $r_0=1$. } \label{f4}
\end{figure}

To linear order in $\delta_i$ and $\phi_i$, the  supercurrent is
\begin{align}\label{eq11}
\mathbf{J}_s=\sum_i \frac{{2e\hbar }}{{{m_i}}}\left| {\Psi _{i0}^2} \right| {\nabla {\phi _i} } - \frac{{4{e^2}}}{c}\sum_i \frac{{\left| {\Psi _{i0}^2} \right|}}{{{m_i}}}{\bf{A}}  \nonumber\\
+\left[ {\frac{{\hbar e}}{{{m_c}}}{\left|\Psi_{10}\Psi_{20}\right|} \left( {\hat x{\partial _x} - \hat y{\partial _y}} \right)\left( {{\delta _2} - {\delta _1}} \right)} \right],
\end{align}
where $\hat{x}$ and $\hat{y}$ are unit vectors in  $x$ and $y$ directions, respectively. The magnetic field is determined by
\begin{align}\label{eq12}
- {\nabla ^2}{\bf{B}} + {\lambda ^{ - 2}}{\bf{B}} = \frac{{8\pi \hbar e{\left|\Psi_{10}\Psi_{20}\right|}}}{{c{m_c}}}\hat z{\partial _{x,y}}\left( {{\delta _1} - {\delta _2}} \right).
\end{align}
We see that $B$ scales linearly with $\delta_i$ and, hence, with the defect potential $\tilde{\alpha}_i(r)$.
For $\tilde{\alpha}_i=\tilde{\alpha}\cos(n\theta)$, the magnetic field ${\bf{B}}\left(\mathbf{r}\right)$ behaves as
\begin{align}\label{eq13}
{\bf{B}}\left(\mathbf{r}\right) \propto \tilde \alpha \left[ {{B_1}\left( r \right)\sin \left[ {\left( {n + 2} \right)\theta } \right] + {B_2}\left( r \right)\sin \left[ {\left( {n - 2} \right)\theta } \right]} \right]\hat z,
\end{align}
The angular dependence is obviously different from that in an $s+is$ state [see Eq. \eqref{eq7}]. The right-hand side of Eq. (\ref{eq12}) is presented in explicit form in Eq. \eqref{eqa10}, and the functions $B_i(r)$ are obtained by substituting \eqref{eqa10} into  (\ref{eq12}) and solving the differential equation.

\section{Numerical calculations}

To complement the analytical analysis we now perform numerical minimization of $\mathcal{F}$ using  time-dependent Ginzburg-Landau equations
\begin{equation}\label{eq5xx}
 \frac{{{\hbar ^2}}}{{2{m_j}D_j}}({\partial _t} + i\frac{{{2e}}}{\hbar }\Phi )\Psi_j   =   - \frac{{\delta \mathcal{F}}}{{\delta {\Psi_j ^*}}},
\end{equation}
\begin{equation}\label{eq6xx}
 \frac{\sigma }{c}(\frac{1}{c}{\partial _t}{\bf{A}} + \nabla \Phi )  =   - \frac{{\delta \mathcal{F}}}{{\delta {\bf{A}}}},
\end{equation}
where $D_j$ is the diffusion constant, $\sigma$ is the normal state conductivity, and $\Phi$ is the electric potential.

We model a defect at a point ${\bf r} =0$ by setting
  $\alpha_i(\mathbf{r})=\alpha_0+\tilde{\alpha}$ for $r<r_0|\cos(n\theta)|$ and $\alpha_i(\mathbf{r})=\alpha_0$ otherwise. The results of the calculation of the induced magnetization for various $n$ are displayed in Fig. \ref{f2}. For $n=0$, there is no spontaneous magnetization in an $s+is$ state,
   while in a $s+id$ state the magnetization is non-zero and displays a two-fold angular variation.  For $n >0$,
    the magnetization in an $s+is$ state displays a $2n$-fold variation, consistent with the result of the linear response theory. In  an $s+id$ state the induced magnetization predominantly displays the  $4$-fold variation for  for $n=2$ and the $2$-fold variation for $n=4$.

     In Fig.  \ref{f3}  we show the amplitude of the induced magnetization $B_z$ as a function of the strength of the defect potential $\tilde{\alpha}$.
        The amplitude  increases linearly with $\tilde{\alpha}$ for an $s+id$ state and quadratically in an $s+is$ state, in full agreement with the analytical results.

         We  also analyzed a square defect with $\alpha_i(\mathbf{r})=\alpha_0+\tilde{\alpha}$ if $-r_0\le x, y \le r_0$ and $\alpha_i(\mathbf{r})=\alpha_0$ otherwise
         (see Fig. \ref{f4}).  The induced magnetization emerges at the corners of the square defect. It has dominant 2-fold (4-fold) angular dependence for an $s+id$ ($s+is$) state. The magnetization pattern in an $s+is$ state has $C_4$ rotation symmetry, while in an $s+id$ state it is invariant under the combination of $C_4$ rotation and time reversal operation.

\section{Relation to  previous work}

Let us compare our results to the previous work. For a circular defect, our results are consistent with those in Ref. \onlinecite{PhysRevLett.102.217002} for an $s+id$ state and in Ref. \onlinecite{PhysRevB.91.161102} for an $s+is$ state. In Ref. \onlinecite{PhysRevB.91.161102} it was argued that, to linear order in defect potential,
for an $s+is$ superconductor {with isotropic impurities, no}  spontaneous supercurrent (and {thus no} induced magnetization) appears unless the $C_4$ symmetry and the time reversal symmetry are both broken. In the present work we extend that result and show that an induced magnetization can also appear in an $s+is$ superconductor even if $C_4$ rotation symmetry is preserved: this requires the defect to break rotational symmetry and perturbative calculations to second order in the defect potential. If the $C_4$ rotation symmetry is explicitly broken by a defect, an induced  magnetization appears already at the  first order in a defect potential.\cite{PhysRevB.91.161102}

\section{Summary}

To summarize, in this paper we have studied the emergence of a  spontaneous magnetization induced by a nonmagnetic defect in a superconductor with either an
  $s+is$ or an $s+id$ pairing symmetry. We found that the  angular dependence of a magnetization depends on the shape of a defect potential and is different for  $s+is$ and  $s+id$ states. For a weak defect, an induced magnetization for an $s+id$ state is linear in  defect potential, while for  an $s+is$ state it is quadratic. For a defect with an $n$-fold angular variation, the magnetization displays a $2n$-fold angular variation for an $s+is$ state and $n\pm2$-fold angular variation
   for an $s+id$ state. Our results show that $s+is$ and $s+id$ pairing symmetries can be distinguished experimentally by measuring  angular dependences of the magnetization patterns.

   The induced magnetization can be measured by imaging method, such as magnetic force microscope and superconducting quantum interference device, and muon-spin relaxation
    experiments.  The shapes of the defects can be controlled by focused ion beam milling. For atomic defects,
      a superconducting order parameter  in a spatially isotropic state is suppressed in a circular region with a radius of the order of  superconducting coherence length.
      As a result, no spontaneous magnetization is induced by atomic defects in an $s+is$ state. Because of this, we believe that the fact that a spontaneous magnetization has not been detected in  zero-field muon-spin relaxation studies of polycrystalline samples  of $\mathrm{Ba_{1-x}K_xFe_2As_2}$ with $0.5\le x \le 0.9$  in Ref. \onlinecite{PhysRevB.89.020502}  does not  actually rule out $s+is$ pairing symmetry.  We hope that future experiments with controlled shape of the defects will be able to resolve the issue whether superconductivity in  $\mathrm{Ba_{1-x}K_xFe_2As_2}$ breaks time-reversal symmetry near $x=1$ and, if yes, whether the superconducting state is $s+is$ or $s+id$.

\begin{acknowledgments}
The authors are indebted to James Sauls and Filip Ronning for helpful discussions. The work by SZL was carried out under the auspices of the U.S. DOE contract No. DE-AC52-06NA25396 through the LDRD program. The work by AVC  was supported by the Office of Basic Energy Sciences, U.S. Department of Energy, under award DE-SC0014402. AVC acknowledges with thanks the hospitality of  the Center for Non-linear Studies, LANL, where he was  2015-2016 Ulam Scholar.

\end{acknowledgments}

\appendix

\section{Calculations of the order parameters}
Here we calculate the variations of the superconducting order parameters in the presence of a weak defect. For the $s+is$ state, the Ginzburg-Landau equations after minimizing  $\mathcal{F}$ in Eq. \eqref{eq1} with respect to $\Psi_i^*$ is
 \begin{align}\label{eqa1}
 \alpha_1\Psi_1+\beta_1|\Psi_1|^2\Psi_1-\frac{\hbar^2}{2m_1}\nabla^2\Psi_1-\frac{\hbar^2}{4m_c}\nabla^2\Psi_2+\gamma_2\Psi_1^*\Psi_2^2=0,
 \end{align}
  \begin{align}\label{eqa2}
 \alpha_2\Psi_2+\beta_2|\Psi_2|^2\Psi_2-\frac{\hbar^2}{2m_2}\nabla^2\Psi_2-\frac{\hbar^2}{4m_c}\nabla^2\Psi_1+\gamma_2\Psi_2^*\Psi_1^2=0,
 \end{align}
where we have set $\gamma_1=\gamma_3=0$ for simplicity. Here we require $\gamma_2>0$ to stabilize the $s+is$ state. We have also neglected the variation of the gauge field by setting $\mathbf{A}=0$ because the fluctuations of $\mathbf{A}$ have the gap of the superconducting gap, while the fluctuations of $\Psi_i$ have a smaller gap in the vicinity of the time reversal symmetry breaking transition. \cite{Lin2012PRL,Stanev11} For a weak defect potential $\alpha_i\rightarrow \alpha_{i}+\tilde{\alpha}_i$ with $|\tilde{\alpha}_i|\ll |\alpha_{i}|$, the change of the order parameter $\Psi_i=\Psi_{i0}(1+\delta_i+i \phi_i)$ is
 \begin{align}\label{eqa3}
 {\alpha _1}\Psi_{10}\left( {{\delta _1} + {\rm{i}}{\phi _1}} \right) + {\beta _1}\left| {\Psi _{10}^2} \right|\Psi_{10}\left( {3{\delta _1} + {{i}}{\phi _1}} \right) \nonumber \\
 - \frac{{{\hbar ^2\Psi_{10}}}}{{2{m_1}}}{\nabla ^2}\left( {{\delta _1} + {{i}}{\phi _1}} \right)
  - \frac{{{\hbar ^2\Psi_{20}}}}{{4{m_c}}}{\nabla ^2}\left( {{\delta _2} + {{i}}{\phi _2}} \right) \nonumber \\
  + \gamma_2\Psi_{10}^* \Psi _{20}^2\left( {{\delta _1} + 2{\delta _2} - {{i}}{\phi _1} + 2{{i}}{\phi _2}} \right) =  - {\tilde \alpha _1},
 \end{align}
and similarly for the second component. By separating the real and imaginary parts, we obtain four linear equations for $\delta_i$ and $\phi_i$. We assume $\tilde{\alpha}_i(\mathbf{r})=\tilde{\alpha}_i(r)\cos(n\theta)$ in the polar coordinate. The Fourier transform is $\tilde{\alpha}_i(k)=\int \tilde{\alpha}_i(r) J_n(k r) r dr$, where $J_n$ is the Bessel function of the first kind. In the momentum space, we need to replace $\nabla^2\rightarrow -k^2$ in Eq. \eqref{eqa3}. By solving Eq. \eqref{eqa3} we find $\delta_i(k)$ and $\phi_i(k)$. By taking the inverse Fourier transform, we obtain $\delta_i(\mathbf{r})=\cos(n\theta)\int\delta_i(k) J_n(kr)k dk$ and similarly for $\phi_i(\mathbf{r})$. For identical two band superconductors with $\alpha_i=\alpha$, $\beta_i=\beta$, $m_i=m$ and $\tilde{\alpha}_i=\tilde{\alpha}$, we have
 \begin{align}
 {\delta _1}\left( k \right) = {\delta _2}\left( k \right) = \frac{{8m\left( {8m\alpha \gamma  + {k^2}\left( { - \beta  + \gamma } \right){\hbar ^2}} \right)m_c^2\tilde \alpha (k)}}{p},
 \end{align}
 \begin{align}
{\phi _1}\left( k \right) =  - {\phi _2}\left( k \right) = \frac{{4{k^2}{m^2}\left( {\beta  - \gamma } \right){\hbar ^2}{m_c}\tilde \alpha (k)}}{p},
 \end{align}
\begin{align}
p\equiv{k^4}{m^2}\left( {\gamma  - \beta } \right){\hbar ^4} \nonumber\\
+ 4\left( {4m\alpha  - {k^2}{\hbar ^2}} \right)\left( {8m\alpha \gamma  + {k^2}\left( {\gamma  - \beta } \right){\hbar ^2}} \right)m_c^2.
\end{align}
The spontaneous magnetization vanishes $\mathbf{B}=0$ according to Eq. \eqref{eq6}. Therefore magnetization appears only in asymmetric two band superconductors.

For the $s+id$ state, we need to replace $\nabla^2\rightarrow \partial_x^2-\partial_y^2$ in the mixed gradient term in Eqs. \eqref{eqa1}, \eqref{eqa2} and \eqref{eqa3}. In the limit $m_c\gg m_i$, we have $\phi_i=0$. For $\alpha_i=\alpha$, $\beta_i=\beta$ and $\tilde{\alpha}_i=\tilde{\alpha}$, but $m_1\neq m_2$, we obtain
\begin{align}
{\delta _1}\left( k \right) = \frac{{2\tilde \alpha (k){m_1}\left( {{k^2}\left( {\beta  - \gamma } \right){\hbar ^2} - 4\alpha \left( {\beta  + \gamma } \right){m_2}} \right)}}{f},
\end{align}
\begin{align}
{\delta _2}\left( k \right) =  - \frac{{2\tilde \alpha (k)\left( {{k^2}\left( { - \beta  + \gamma } \right){\hbar ^2} + 4\alpha \left( {\beta  + \gamma } \right){m_1}} \right){m_2}}}{f},
\end{align}
\begin{align}
f\equiv {k^4}\left( { - \beta  + \gamma } \right){\hbar ^4} + 4{k^2}\alpha \beta {\hbar ^2}{m_2} \nonumber\\
+ 4\alpha {m_1}\left( {{k^2}\beta {\hbar ^2} - 4\alpha \left( {\beta  + \gamma } \right){m_2}} \right).
\end{align}
Denote $\delta_1(\mathbf{r})-\delta_2(\mathbf{r})=g(r)\cos(n\theta)$, we have
\begin{align}\label{eqa10}
{\partial _{x,y}}\left( {{\delta _1} - {\delta _2}} \right)=n\cos(2\theta )\sin(n\theta )\left( g - r\partial_r g \right) \nonumber\\
+ \frac{\sin(2 \theta) \cos( n\theta )}{2}\left( n^2 g + r\left(  - \partial_r g + r\partial_{r}^2 g\right) \right).
\end{align}
From Eq. \eqref{eq12}, we obtain the angular dependence of $\mathbf{B}$ in Eq. \eqref{eq13}.

\bibliography{reference}

\end{document}